\documentclass[12pt,a4paper]{article}
\usepackage[latin1]{inputenc}
\usepackage[T1]{fontenc}
\usepackage{amsmath}
\usepackage{amsfonts}
\usepackage{amssymb}
\usepackage[english]{babel}
\usepackage[dvips]{graphicx,color}
\title{A new path integral representation for the thermal partition
function}
\author{C.~D.~Fosco$^{a}$\\
A.~P.~C.~Malbouisson$^{b}$\\ and I.~Roditi$^{b}$\\ {\normalsize\it
$^a$Centro At\'omico Bariloche and Instituto Balseiro}\\
{\normalsize\it Comisi\'on Nacional de Energ\'\i a At\'omica} \\
{\normalsize\it 8400 Bariloche, Argentina.}\\ {\normalsize\it
$^b$Centro Brasileiro de Pesquisas F\'isicas - CBPF/MCT}\\
{\normalsize\it Rua Dr. Xavier Sigaud, 150, 22290-180 Rio de Janeiro,
RJ, Brazil}}
\begin{document}
\newcommand*{\be}{\begin{equation}} \newcommand*{\ee}{\end{equation}}
\newcommand*{\beq}{\begin{eqnarray}} \newcommand*{\eeq}{\end{eqnarray}}
\def\d{{\rm d}}
\maketitle
% ====================================================================
\begin{abstract}
  \noindent The boundary conditions corresponding to the Matsubara
formalism for the $T > 0$ partition function may be introduced as {\em
constraints\/} in the path integral for the vacuum amplitude. We implement
those constraints with time-independent Lagrange multipliers and, by
integrating out the original fields, we obtain an alternative
representation for the partition function, in terms of the Lagrange
multipliers as dynamical fields.  The resulting functional integral has the
appealing property of involving only $d$-dimensional, {\em time
independent\/} fields, and looks like a nonlocal version of the classical
partition function.  We develop this formalism within the context of the
scalar and Dirac fields.
\end{abstract}
% ====================================================================
%%%%%%%%%%%%%%%%%%%%%%%%%%%%%%%%%%%%%%%%%%%%%%%%%%%%%%%%%%%%%%%%%%%%%
%%%%%%%%%%%%%%%%%%%%%%%%%%%%%%%%%%%%%%%%%%%%%%%%%%%%%%%%%%%%%%%%%%%%%
%%%%%%%%%%%%%%%%%%%%%%%%% Introduction %%%%%%%%%%%%%%%%%%%%%%%%%%%%%%
%%%%%%%%%%%%%%%%%%%%%%%%%%%%%%%%%%%%%%%%%%%%%%%%%%%%%%%%%%%%%%%%%%%%%
%%%%%%%%%%%%%%%%%%%%%%%%%%%%%%%%%%%%%%%%%%%%%%%%%%%%%%%%%%%%%%%%%%%%%
Many years ago~\cite{mats1}, the theoretical foundations for a systematic
treatment of Quantum Field Theory (QFT) at finite temperature ($T>0$) and
its relativistic generalization~\cite{ume1} were laid down. That approach,
the now called Matsubara (or imaginary time) formalism has been very
successful indeed in allowing for the evaluation of thermal effects in QFT,
both in the High Energy~\cite{kap1} and Condensed Matter Physics realms.
Besides, the notion of Abelian and non Abelian gauge fields, with all its
consequences for particle physics~\cite{kap1,le1,belac1} could also be set
up in the $T > 0$ context in a quite natural way.

A fundamental property introduced by this formalism is the imaginary-time
periodicity (antiperiodicity) conditions for the bosonic (fermionic)
fields, something that may be already discovered at the level of the
partition function, ${\mathcal Z}(\beta)$, for a system at a temperature $T
= 1/\beta$:
\begin{equation} 
{\mathcal Z}(\beta) \;=\; {\rm Tr} \big[ e^{-\beta {\hat
H}} \big]\,.
\end{equation} 
Assuming, for the sake of simplicity, that the system has only one degree
of freedom, $q(\tau)$, the expression above may be written more explicitly
as follows: 
\begin{equation} 
{\mathcal Z}(\beta) \;=\; \int dq \; \langle q| e^{-\beta
H}|q\rangle\;=\; \int dq \; \langle q,\beta | q , 0 \rangle 
\end{equation}
where we have introduced the rotating basis $|q,\tau\rangle$, for a {\em
fictitious\/} imaginary time evolution.  Then the standard path integral
construction for the transition amplitude between the times $\tau=0$ and
$\tau=\beta$ may be applied, to write the partition function in the
Matsubara formalism: 
\begin{equation}\label{eq:defzh} 
{\mathcal Z}(\beta) \;=\;
  \int_{\rm q(0)=q(\beta)}\, {\mathcal D} p \, {\mathcal D} q \; 
  e^{\int_0^\beta d\tau \big[ i p \dot{q} 
    \,-\, H(p,q)\big]} \;,
\end{equation}
where one only considers phase-space paths $q(\tau), \, p(\tau)$ ($\tau \in
[0,\beta]$) such that $q(0) = q(\beta)$.  There is no need to impose any
boundary condition on the $p(\tau)$ paths~\footnote{Albeit a more symmetric
form for the boundary conditions in the path integral for ${\mathcal
Z}_0(\beta)$ can be used~\cite{teitelboim}.}.

In the usual case, one may integrate out the canonical momentum variables,
what leads, for the case of a $d+1$-dimensional field theory, to a model
defined on $S^1 \times R^{d}$, where the radius of $S^{1}$ is proportional
to the inverse temperature $\beta$. In Fourier space, the corresponding
frequencies are the usual discrete Matsubara frequencies.

A characteristic feature of the Matsubara formalism (shared by the
real-time formulation) is that the introduction of a time dependence for
the fields is seemingly unavoidable, even when one limits oneself to the
calculation of {\em static\/} thermal properties.

With the aim of constructing a new representation where only static fields
are involved, in this letter, we introduce an alternative way of dealing
with the $T>0$ QFT partition function. It is inspired by a recent paper in
which a constrained functional integral approach is used to implement the
effect of fluctuating boundaries in the Casimir effect~\cite{kardar}.  In
the present context, this allows one to introduce the periodicity
conditions by means of Lagrange multipliers ($d$-dimensional when the field
lives in $d+1$ dimensions).  Then the original fields can be integrated,
what leaves a functional depending only on the $d$-dimensional Lagrange
multipliers.

% ====================================================================
%%%%%%%%%%%%%%%%%%%%%%%%%%%%%%%%%%%%%%%%%%%%%%%%%%%%%%%%%%%%%%%%%%%%%
%%%%%%%%%%%%%%%%%%%%%%%%%%%%%%%%%%%%%%%%%%%%%%%%%%%%%%%%%%%%%%%%%%%%%
%%%%%%%%%%%%%%%%%%%%%%%%%% Construction %%%%%%%%%%%%%%%%%%%%%%%%%%%%%
%%%%%%%%%%%%%%%%%%%%%%%%%%%%%%%%%%%%%%%%%%%%%%%%%%%%%%%%%%%%%%%%%%%%%
%%%%%%%%%%%%%%%%%%%%%%%%%%%%%%%%%%%%%%%%%%%%%%%%%%%%%%%%%%%%%%%%%%%%%
We shall introduce the new method by considering first the case of a single
degree of freedom, extending it to more complex systems afterwards. We
begin by writing the phase-space path integral for ${\mathcal Z}_0$, the
{\em vacuum\/} functional for a single quantum degree of freedom:
\begin{equation}\label{eq:defz0} 
{\mathcal Z}_0 \;=\; \int \, {\mathcal D}p
  \,{\mathcal D}q \;
  \exp \big\{-S_0[q(\tau),p(\tau)] \big\} \;,
\end{equation}
where $S_0$ is the first-order action, \mbox{$S_0 =\int_{-\infty}^{+\infty}
  d\tau \,{\mathcal L}$}, with ${\mathcal L} = - i p \dot{q} + {\mathcal
  H}(p,q)$, with ${\mathcal H}$ denoting the Hamiltonian, assumed (for the
sake of simplicity) to be of the form: ${\mathcal H}(p,q)=T(p)+V(q)$.

Of course, ${\mathcal Z}_0$ may be regarded as the limit of an
imaginary-time
transition amplitude,
\begin{eqnarray}
  {\mathcal Z}_0 &=& \lim_{T \to +\infty} \,
  \langle q_0, T | q_0, -T \rangle  \nonumber\\
  &=& \lim_{T \to +\infty} \,\sum_n | \langle q_0| n \rangle |^2 e^{-2 T
E_n}
  \; = \;\lim_{T \to +\infty} \, | \langle q_0| 0 \rangle |^2 e^{-2 T E_0}
\end{eqnarray}
where we have introduced $|n\rangle$, the eigenstates of ${\hat H}$, ${\hat
H}|n\rangle = E_n |n\rangle$, and $q_0$, the asymptotic value for $q_0$ at
$T \to \pm \infty$ (usually, $q_0 \equiv 0$). $E_0$ is the energy of
$|0\rangle$, the ground state.

We have already recalled the usual functional integral for the thermal
partition function ${\mathcal Z}_0(\beta)$ in the Matsubara formalism. Let
us now show how one can write an alternative expression for the same
object, starting from the vacuum transition amplitude, ${\mathcal Z}_0$,
and imposing a periodicity constraint on the paths.  We first use the
superposition principle, introducing decompositions of the identity at
$\tau = 0$ and $\tau = \beta$, in order to write ${\mathcal Z}_0$ in an
equivalent way: 
\begin{equation} 
{\mathcal Z}_0 \;=\; \lim_{T\to \infty}
\,\int dq_2 dq_1 \,
  \langle q_0, T | q_2, \beta \rangle \, \langle q_2, \beta | q_1 , 0
\rangle
  \, \langle q_1, 0 | q_0, -T \rangle \;,
\end{equation}
or, in a path integral representation,
\begin{eqnarray}\label{eq:aux1} 
{\mathcal Z}_0 &=& \lim_{T \to \infty} \;
  \int dq_2 dq_1 \, \int_{q(\beta)= q_2}^{q(T) = q_0} \, 
  {\mathcal D}p \,{\mathcal D}q \; e^{- \int_\beta^T d\tau {\mathcal L}}
  \nonumber\\
  &\times& \int_{q(0)=q_1}^{q(\beta)=q_2} \, {\mathcal D}p \,{\mathcal
D}q\;
  e^{- \int_0^\beta d\tau {\mathcal L}}
  \;\int_{q(-T)=q_0}^{q(0)=q_1} \, {\mathcal D}p \,{\mathcal D}q\;
  e^{- \int_{-T}^0 d\tau {\mathcal L}} \;.
\end{eqnarray}
The representation above is useful in order to show that the naive
procedure of introducing a periodicity constraint for just the coordinate
$q(\tau)$ is not the right procedure to obtain ${\mathcal Z}(\beta)$ from
the path integral for ${\mathcal Z}_0$. Indeed, we can see that
$$
\int \, {\mathcal D}p \,{\mathcal D}q \; \delta\big( q(\beta) - q(0)\big)
\exp
\big\{-S_0[q(\tau),p(\tau)] \big\}
$$
\begin{equation}
  \; = \; \lim_{T\to \infty} \,e^{ - E_0 ( 2 T - \beta)} \,  
  \int dq_1 |\langle 0 | q_1 \rangle |^2 \,
  \langle q_1 , \beta | q_1 , 0 \rangle \;,
\end{equation}
and taking the ratio with the (unconstrained) vacuum functional,
\begin{equation}
 \frac{\int \,{\mathcal D}p \,{\mathcal D}q \;\delta\big(
q(\beta) - q(0)\big)\, e^{-S_0}}{\int \,{\mathcal D}p\,{\mathcal D}q \,
e^{-S_0}} \, = \, e^{ \beta E_0 } \, \int dq_1 |\langle 0 | q_1 \rangle |^2
\, \langle q_1 , \beta | q_1 , 0 \rangle \;, 
\end{equation} 
where it is evident that we cannot extract a ${\mathcal Z}(\beta)$ factor.
There is, however, a simple alternative that does the job: to reproduce
${\mathcal Z}_0(\beta)$ we impose periodicity conditions for both phase
space variables, i.e., we introduce the object, ${\mathcal Z}_s(\beta)$
defined
as
follows:
\begin{equation} 
{\mathcal Z}_s(\beta) \;\equiv \; \frac{\int \,{\mathcal
D}p
    \,{\mathcal D}q \;\delta\big( q(\beta) - q(0)\big)\,
\delta\big(p(\beta) -
    p(0)\big) \, e^{-S_0}}{\int \,{\mathcal D}p\,{\mathcal D}q \, e^{-S_0}}
  \,.
\end{equation}
Using the superposition principle, we can see that:
$$
\int \, {\mathcal D}p \,{\mathcal D}q \; \delta\big( q(\beta) - q(0)\big)
\delta\big( p(\beta) - p(0)\big) \; e^{-S_0}= \lim_{T\to \infty} \,\int
dp_1
dq_1 \,\Big[
$$
\begin{equation}
  \langle q_0, T | p_1, \beta  \rangle \,
  \langle p_1, \beta | q_1, \beta \rangle \,
  \langle q_1 , \beta | q_1 , 0 \rangle \,
  \langle q_1 , 0 | p_1 , 0 \rangle\,
  \langle p_1 , 0 | q_0 , -T \rangle \Big]
\end{equation}
or
$$
\int \, {\mathcal D}p \,{\mathcal D}q \; \delta\big( q(\beta) - q(0)\big)
\delta\big( p(\beta) - p(0)\big) e^{-S_0} \,=\, \lim_{T\to \infty} \,e^{ -
E_0
  ( 2 T - \beta)}
$$
$$
\times \, \int \frac{dp_1 dq_1}{2\pi} \, \langle q_0|0\rangle \langle 0 |
p_1\rangle \, \langle q_1 , \beta | q_1 , 0 \rangle \, \langle p_1|0\rangle
\langle 0 | q_0\rangle
$$
\begin{equation}
  = \, \lim_{T\to \infty} \,e^{ - E_0 ( 2 T - \beta)}
  \,|\langle q_0 | 0 \rangle |^2 \,\int dq_1 \, \langle q_1 , \beta | q_1 ,
0
  \rangle \;.
\end{equation}
We have used the property
\begin{equation}
  \langle p_1, \beta | q_1, \beta \rangle \; 
  \langle q_1 , 0 | p_1 , 0 \rangle \;=\;\langle p_1| q_1\rangle
  \langle q_1| p_1\rangle = \frac{1}{2\pi} \;.
\end{equation}
Then we conclude that
\begin{equation} 
{\mathcal Z}_s(\beta) \;=\; {\rm Tr} \big[ e^{-\beta \,
    :{\hat H}:} \big]
\end{equation}
where $:{\hat H}:$ denotes the normal-ordered Hamiltonian operator, i.e.:
\begin{equation}
  :{\hat H}: \;\equiv\;   {\hat H} \, - \, E_0  \,.
\end{equation}
The conclusion of the previous derivation is then that, by imposing
periodicity in both phase space variables, and discarding
$\beta$-independent
factors (as required by the normalization) we obtain ${\mathcal
Z}_s(\beta)$,
the partition function corresponding to the original Hamiltonian with
energies
redefined to $0$ for the ground state. The subtraction of the vacuum energy
is
usually irrelevant (except for some exceptional situations), as it is wiped
out when taking derivatives of the free energy to calculate physical
quantities.

% ====================================================================
%%%%%%%%%%%%%%%%%%%%%%%%%%%%%%%%%%%%%%%%%%%%%%%%%%%%%%%%%%%%%%%%%%%%%
%%%%%%%%%%%%%%%%%%%%%%%%%%%%%%%%%%%%%%%%%%%%%%%%%%%%%%%%%%%%%%%%%%%%%
%%%%%%%%%%%%%%%%%%%%%% Harmonic Oscillator %%%%%%%%%%%%%%%%%%%%%%%%%%
%%%%%%%%%%%%%%%%%%%%%%%%%%%%%%%%%%%%%%%%%%%%%%%%%%%%%%%%%%%%%%%%%%%%%
%%%%%%%%%%%%%%%%%%%%%%%%%%%%%%%%%%%%%%%%%%%%%%%%%%%%%%%%%%%%%%%%%%%%%
As a test, let us apply the previous results to the case of the harmonic
oscillator in one spatial dimension.  To proceed, we exponentiate the two
$\delta$-functions, by means of Lagrange multipliers $\xi_1$ and $\xi_2$.
They
are just real variables in this case, and we are thus lead to a new
expression
for ${\mathcal Z}_s(\beta)$, which may be written as follows:
\begin{eqnarray}\label{eq:aux2} 
{\mathcal Z}_s(\beta) &=& {\mathcal N} \,
\int
  {\mathcal D} p \, {\mathcal D} q
  \int_{-\infty}^{\infty} \frac{d\xi_1}{2\pi} \int_{-\infty}^{\infty}
  \frac{d\xi_2}{2\pi} \nonumber\\
  &\times& e^{i\int_{\infty}^{+\infty} d\tau \Big[
    q(\tau) \, \xi_1 \,\big( \delta(\tau-\beta) - \delta(\tau) \big)
    \,+\, p(\tau)\, \xi_2 \, \big( \delta(\tau-\beta) - \delta(\tau) \big)}
  \nonumber\\
  &\times& e^{\int_{-\infty}^\infty d\tau \big[ i p \dot{q} 
    \,-\,\frac{1}{2} (p^2 + \omega^2 q^2 )\big]} \;,
\end{eqnarray}
where the constant ${\mathcal N}$ is used to denote the $({\mathcal
  Z}_0)^{-1}$ factor.

The integral over $p$ and $q$ (which we shall perform first) is obviously
Gaussian. Indeed, interchanging the order of integration, we see that
(\ref{eq:aux1}) may be rewritten as follows:
\begin{eqnarray}\label{eq:zb2} 
{\mathcal Z}_s(\beta) &=& {\mathcal N}
  \,\int_{-\infty}^{\infty}
  \frac{d\xi_1}{2\pi} \int_{-\infty}^{\infty}\frac{d\xi_2}{2\pi}
  \int {\mathcal D} Q 
  \nonumber\\
  &\times& e^{- \frac{1}{2} \int_{\infty}^{+\infty} d\tau \, Q_a(\tau) 
    \widehat{{\mathcal K}}_{ab} Q_a(\tau) \,+\,i\, \int_{-\infty}^\infty
d\tau
    j_a(\tau) Q_a(\tau) } \;, 
\end{eqnarray}
where we have introduced \mbox{$Q \,=\, (Q_a)$} ($a=1,\, 2$) for the
canonical
coordinates, such that $Q_1 \equiv q$ and $Q_2 \equiv p$. Also,
\mbox{$j_a(\tau) \equiv \xi_a \,( \delta(\tau-\beta) - \delta(\tau))$}, and
$\widehat{\mathcal K}_{ab}$ are the elements of the $2 \times 2$ operator
matrix $\widehat{\mathcal K}$, given by:
\begin{equation}\label{eq:defk}
  \widehat{\mathcal K} \;=\; 
  \left( 
    \begin{array}{cc}
      \omega^2              & i \frac{d}{d\tau} \\
      - i \frac{d}{d\tau}     & 1 
    \end{array}
  \right) \;.
\end{equation}
The result of integrating out $Q_a$, may be written as follows:
\begin{equation} 
{\mathcal Z}_s(\beta) \;=\; {\mathcal N}\, 2\pi \,
  \big(\det\widehat{\mathcal K}\big)^{-\frac{1}{2}} \; \int
  \frac{d^2\xi}{(2\pi)^2} \,\, e^{- \frac{1}{2} \xi_a M_{ab} \xi_b } \;,
\end{equation}
with
\begin{equation}\label{eq:defm}
  M \;\equiv\; 2 \, \Omega(0) \,-\, \Omega(\beta) \,-\, \Omega(-\beta) \;, 
\end{equation}
where $\Omega(\tau)$ denotes the inverse of the operator ${\mathcal K}$ of
(\ref{eq:defk}); its explicit form may be easily found to be:
\begin{equation}\label{eq:defom}
  \Omega(\tau)\;\equiv\;
  \left( 
    \begin{array}{cc}
      \frac{1}{2\omega} & \frac{i}{2} {\rm sgn}(\tau) \\
      -\frac{i}{2} {\rm sgn}(\tau) & \frac{\omega}{2} 
    \end{array}
  \right) \; e^{-\omega |\tau|} 
\end{equation}
(${\rm sgn} \equiv$ sign function). Equation (\ref{eq:defom}) can be used
in
(\ref{eq:defm}), to write:
\begin{equation}\label{eq:m1}
  M\;=\;
  \left( 
    \begin{array}{cc}
      \omega^{-1} & 0 \\
      0 & \omega 
    \end{array}
  \right) \; (n_B(\omega) + 1)^{-1} \;,
\end{equation}
where
\begin{equation}
  n_B(\omega) \equiv (e^{\beta \omega} - 1)^{-1}\
\end{equation}
is the Bose-Einstein distribution function (with the zero of energy set at
the
ground state).

The resulting integral over the $\xi_a$ variables becomes:
\begin{equation}\label{eq:zosc} 
{\mathcal Z}_s(\beta) \;=\;
  \int \frac{d^2\xi}{2\pi} \,
  e^{- \frac{\omega^{-1} \, \xi_1^2 \,+\; 
      \omega \; \xi_2^2}{2 [ n_B(\omega) + 1 ]}} \;.
\end{equation}
This integral is over two real variables $\xi_a$, which are $0$-dimensional
fields, one dimension less than the $0+1$ dimensional original theory. This
integral may be compared with the one corresponding to classical
statistical
mechanics. To that end, we evaluate the partition function in the classical
(high-temperature) limit.  In that limit, we approximate the integrand and
${\mathcal Z}_s(\beta)$ becomes:
\begin{equation}\label{eq:zosc2} 
{\mathcal Z}_s(\beta) \;\simeq\;\int
  \frac{d^2\xi}{2\pi} \,
  e^{- \beta H(\xi_1,\xi_2)} \;\;\; (\beta << 1) \;,
\end{equation}
where:
\begin{equation}
  H(\xi_1,\xi_2) \;\equiv\; \frac{1}{2} \big( \xi_1^2 \,+\, \omega^2
\xi_2^2 
  \big)\;.
\end{equation}
We see that (\ref{eq:zosc2}) corresponds exactly to the classical partition
function for a harmonic oscillator, when the identifications: $\xi_1=p$
(classical momentum), and $\xi_2=q$ (classical coordinate) are made
\begin{equation}\label{eq:zosc3} {\mathcal Z}_s(\beta) \;\simeq\;\int
\frac{dp
    dq}{2\pi} \,
  e^{- \beta \frac{1}{2} \big( p^2 \,+\, \omega^2 q^2\big)}
  \;\;\; (\beta << 1) \;.
\end{equation}

On the other hand, had the exact form of the integral been kept (no
approximation), we could still have written an expression similar to the
classical partition function, albeit with an `effective Hamiltonian'
$H_{eff}(\xi_1,\xi_2)$:
\begin{equation}\label{eq:zosc4} 
{\mathcal Z}_s(\beta) \;=\;\int
  \frac{d^2\xi}{2\pi} \,
  e^{- \beta H_{eff}(\xi_1,\xi_2)} \;,
\end{equation}
where:
\begin{equation}
  H_{eff}(\xi_1,\xi_2) \;\equiv\; \frac{1}{2 \beta} \,
  \big( n_B(\omega) + 1 \big)^{-1}\, \big(\omega^{-1} \, \xi_1^2 \,+\; 
  \omega \; \xi_2^2 \big) \;.
\end{equation}
This shows that the quantum partition function may also be written as a
classical one, by using a $\beta$-dependent Hamiltonian, which of course
tends
to its classical counterpart in the high-temperature limit.  This
representation is also valid for interacting theories, see below.

By integrating out the Lagrange multipliers in the (exact) expression for
the
partition function (\ref{eq:zosc}), we obtain:
\begin{equation}\label{eq:free} 
{\mathcal Z}_s(\beta) \;=\; n_B(\omega)
\,+\,
  1 \;=\;
  \frac{1}{1\,-\, e^{-\beta \omega}} \;.
\end{equation}
which is the correct result. In what follows, to simplify the notation, we
shall omit writing the `$s$' subscript in ${\mathcal Z}(\beta)$, assuming
implicitly that one is dealing with the normal-ordered Hamiltonian.

An important fact that has emerged from an analysis of the classical
(high-temperature) limit: the Lagrange multipliers have a physical
interpretation.  The multiplier associated to the periodicity condition for
$q$ plays the role of a classical momentum, while the one corresponding to
the
periodicity for the momentum becomes a generalization of the classical
coordinate. The same interpretation might also be retained far from the
classical limit, but then the Hamiltonian departs from the classical one,
receiving quantum corrections.

The extension to the QFT of a real scalar field $\varphi$ in $d+1$
(Euclidean)
dimensions is quite straightforward. Let $\varphi(x)=\varphi(\tau,{\mathbf
  x})$ where $x=(\tau,{\mathbf x}) \in {\mathbb R}^{(d+1)}$, $\tau \in
{\mathbb R}$ and ${\mathbf x} \in {\mathbb R}^{(d)}$.  The free Euclidean
action, $S_0$, defined in terms of the phase-space variables is
\begin{equation}\label{eq:defs0}
  S_0 \,=\,\int d^{d+1}x \,\Big[ - i \pi \partial_\tau \varphi 
  + {\mathcal H}_0(\pi,\varphi) \Big] \;,
\end{equation}
with
\begin{equation} {\mathcal H}_0(\pi,\varphi) \;\equiv\; \frac{1}{2} \Big[
  \pi^2 \,+\, |{\mathbf \nabla}\varphi |^2 + m^2 \varphi^2 \Big] \;.
\end{equation}

We then have to implement the periodic boundary conditions both for
$\varphi(\tau,{\mathbf x})$ and its canonical momentum $\pi(\tau,{\mathbf
x})$
\begin{equation}
  \varphi\left(\beta,{\mathbf x}\right) \,=\, \varphi \left(0,{\mathbf
      x}\right)\;,\;\;\; \pi\left(\beta,{\mathbf x}\right) \,=\,
\pi\left(0,{\mathbf
      x}\right)\;,\;\;\forall\, {\mathbf x} \, \in {\mathbb R}^{(d)} \;,
\end{equation}
which requires the introduction of two {\em static\/} Lagrange multiplier
fields: $\xi_a({\mathbf x})$, $a=1,\,2$.  Defining the two-component field
\mbox{$\Phi = (\Phi_a)$}, \mbox{$a=1,\,2$}, such that $\Phi_1 = \varphi$
and
$\Phi_2 = \pi$, an analogous procedure to the one followed for the harmonic
oscillator yields:
\begin{equation}\label{eq:zesc1} 
{\mathcal Z}_0(\beta) \;=\; {\mathcal N}
  \,\int \,
  {\mathcal D}\xi \, \int {\mathcal D}\Phi \; 
  e^{-\frac{1}{2} \int d^{d+1}x \, \Phi_a 
    {\hat{\mathcal K}}_{ab} \Phi_b \,+\,i\, \int d^{d+1}x  j_a \Phi_a}
  \;,
\end{equation}
where $j_a(x) \equiv \xi_a({\mathbf x}) \big( \delta(\tau-\beta) -
\delta(\tau) \big)$ and:
\begin{equation}\label{eq:newdefk} 
{\widehat{\mathcal K}}\;=\;
  \left( 
    \begin{array}{cc}
      {\hat h}^2       & i \frac{\partial}{\partial\tau} \\
      - i \frac{\partial}{\partial\tau}     & 1 
    \end{array}
  \right) \;,
\end{equation}
where we have introduced \mbox{${\hat h} \equiv \sqrt{-\nabla^2 + m^2}$},
the
first-quantized energy operator for massive scalar particles.

Performing the integral over $\Phi$, yields the partition function in terms
of
the Lagrange multipliers:
\begin{equation} 
{\mathcal Z}_0(\beta) \;=\; \int {\mathcal D}\xi \,e^{-
    \frac{1}{2} \int d^dx \int d^dy \, \xi_a ({\mathbf x}) \; \langle
{\mathbf
      x} | {\hat M}_{ab}| {\mathbf y} \rangle \; \xi_b ({\mathbf y})} \;,
\end{equation}
with ${\hat M}\equiv 2 {\hat\Omega}(0) - {\hat \Omega}(\beta) -
{\hat\Omega}(-\beta)$ and
\begin{equation}\label{eq:newdefom} {\hat \Omega}(\tau)\;\equiv\;
  \left( 
    \begin{array}{cc}
      \frac{1}{2} {\hat h}^{-1} 
      & \frac{i}{2} {\rm sgn}(\tau) \\
      -\frac{i}{2} {\rm sgn}(\tau) 
      & \frac{1}{2} {\hat h}
    \end{array}
  \right) \; e^{-{\hat h} |\tau|}  \;.
\end{equation}
Then,
\begin{equation}\label{eq:newm} 
{\hat M}\;\equiv\;
  \left( 
    \begin{array}{cc}
      {\hat h}^{-1} 
      & 0 \\
      0 
      & {\hat h}
    \end{array}
  \right) \;({\hat n}_B + 1)^{-1}\;,
\end{equation}
where
\begin{equation} 
{\hat n}_B \;\equiv\; \frac{1}{e^{\beta {\hat h}} - 1} \;.
\end{equation}
Coming back to the expression for ${\mathcal Z}_0(\beta)$, we see that:
\begin{eqnarray}\label{eq:zbose0} 
{\mathcal Z}_0(\beta) &=&
  \int {\mathcal D}\xi \,\exp \Big\{ 
  - \frac{1}{2} \int d^dx \int d^dy \big[ \xi_1 ({\mathbf x}) \;
  \langle {\mathbf x} |{\hat h}^{-1} \, ({\hat n}_B+1)^{-1} | {\mathbf y}
\rangle \;
  \xi_1
  ({\mathbf y}) \nonumber\\
  &+& \xi_2({\mathbf x}) \;\langle {\mathbf x} |{\hat h} \, ({\hat
n}_B+1)^{-1} |
  {\mathbf y} \rangle \; 
  \xi_2 ({\mathbf y}) \big] \Big\} \;.
\end{eqnarray}
By a simple field redefinition, we see that:
\begin{equation} 
{\mathcal Z}_0(\beta) \;=\; \det \big( {\hat n}_B + 1\big)
\end{equation}
which can be evaluated in the basis of eigenstates of momentum to yield:
\begin{equation} 
{\mathcal Z}_0(\beta) \;=\; \prod_{\mathbf k} \big[
  n_B(E_{\mathbf k}) + 1 \big]
\end{equation}
where $E_{\mathbf k} \equiv \sqrt{{\mathbf k}^2 + m^2}$. The free-energy
density, $F_0(\beta)$, is of course:
\begin{equation}
  F_0(\beta)\;=\; \frac{1}{\beta} \, \int \frac{d^dk}{(2\pi)^d} \, \ln
\big( 1
  \,-\, e^{-\beta E_{\mathbf k}}\big) \;.
\end{equation}

In the classical, high-temperature limit, the path integral for the
partition
function becomes:
\begin{equation} 
{\mathcal Z}_0(\beta) \;\simeq\; \int {\mathcal D}\xi \,
e^{-\beta \; H(\xi)} \;,
\end{equation}
where:
\begin{equation}
  H(\xi) \;=\;\frac{1}{2} \int d^dx \big[ \xi_1^2({\mathbf x}) \,+\,
  |{\mathbf\nabla}\xi_2({\mathbf x})|^2 \,+\, m^2 \, \xi_2^2({\mathbf x}) 
  \Big] \;.
\end{equation}
This is, again, the usual classical expression for the partition function,
with the Lagrange multipliers playing the role of phase space variables,
and the integration measure being the corresponding Liouville measure. 
Besides, it is clear that the representation (\ref{eq:zbose0}) always
involves static fields, unlike in the Matusbara formalism. The price to pay
for this `dimensional reduction' is that the resulting `action' (the
exponent of the functional to be integrated) is spatially non local. It
becomes local only in the high-temperature limit.

When self-interactions are included, instead of the free action $S_0$, we
must consider instead
\begin{equation}
  S\;=\;S_0 \,+\, S_I \;,
\end{equation}
where we assume that $S_I = \int d^{d+1}x \,V(\varphi)$, $V(\varphi)$ being
an
even polynomial in $\varphi$. The new representation for the partition
function is then:
\begin{equation}\label{eq:zescint1} 
{\mathcal Z}(\beta) \;=\; {\mathcal N}  \,\int \,
  {\mathcal D}\xi \, \int {\mathcal D}\Phi \; 
  e^{- S (\Phi) \,+\,i\, \int d^{d+1}x  j_a \Phi_a}
  \;.
\end{equation}

Recalling the definition of the generating functional for $T=0$ connected
correlation functions, ${\mathcal W}$, we may write:
\begin{equation}\label{eq:defw} 
{\mathcal N} \, \int {\cal D}\Phi\;
  \exp\left[- S(\Phi) + i \int \,
    d^{d+1}x j_a \Phi_a \right]\; \equiv \; e^{-{\mathcal W}(j)}\,,
\end{equation}
so that the partition function ${\mathcal Z}(\beta)$ becomes:
\begin{equation}\label{eq:Wxi} 
{\mathcal Z} (\beta) \;=\; \int {\cal
D}\xi\;
  e^{- {\mathcal W}(j)}\;,
\end{equation}
a formula that makes contact with $T=0$ QFT, building the thermal
corrections
on top of the $T=0$ ones. We have use the small $j$ to denote the
$2$-component current which is a function of the Lagrange multipliers. A
capital $J$ shall be reserved to denote a completely arbitrary
$2$-component
source, so that:
\begin{equation}\label{eq:defw1} 
{\mathcal N} \, \int {\cal D}\Phi\;
  exp\left[- S(\Phi) + i \int \,
    d^{d+1}x J_a \Phi_a \right]\; \equiv \; e^{-{\mathcal W(J)}}\,.
\end{equation}

Defining now $H_{eff}(\xi)$, the `effective Hamiltonian' for ${\xi}$, by
means of the expression:
\begin{equation}
  H_{eff}(\xi) \;=\; \frac{1}{\beta} \, {\mathcal W}(j) \;,
\end{equation}
we see that the partition function is simply given by:
\begin{equation} 
{\mathcal Z}(\beta)\; = \; \int {\mathcal D}\xi \;
  \exp\left[- \beta H_{eff}(\xi) \right] \;,
\end{equation}
one of the most important results in this article, since it yields the path
integral for the quantum partition function as a classical-looking
functional integral, involving an effective Hamiltonian which takes into
account all the $T=0$ quantum effects. Indeed, the usual functional
expansion for ${\mathcal W}(J)$ is:
\begin{eqnarray}\label{eq:Wexpansion} 
{\mathcal W}(J) &=&
  \sum_{n=2}^{\infty}\frac{1}{n!}\int d^{d+1}x_1\cdots
  d^{d+1}x_n\, \nonumber\\
  &\times& {\mathcal W}^{(n)}_{a_1 \ldots
    a_n}(x_1,\,\ldots,\,x_n)\,J_{a_1}(x_1)\cdots J_{a_n}(x_n),
\end{eqnarray}
where ${\mathcal W}^{(n)}$ is $i^n$ times the $n$-point connected
correlation function for the field $\Phi$. Thus, knowing those correlation
functions, at a given order in a loop expansion, one may obtain an
analogous functional expansion for the effective Hamiltonian. Indeed, to do
that, one should perform the integrations over the imaginary-time variables
(taking advantage of the $\delta$-functions):
\begin{eqnarray}\label{eq:Heffexpansion}
  H_{eff}(\xi) &=& \sum_{n=2}^{\infty}\frac{1}{n!}\int
  d^d{\mathbf x}_1 \cdots d^d{\mathbf x}_n\,H^{(n)}_{a_1 \ldots
    a_n}({\mathbf x}_1,\,\ldots,\,{\mathbf x}_n)\nonumber\\
  &\times& \xi_{a_1}({\mathbf x}_1) \cdots \xi_{a_n}({\mathbf x}_n) \;,
\end{eqnarray}
where
\begin{eqnarray}
  H^{(n)}_{a_1 \ldots a_n}({\mathbf x}_1,\,\ldots,\,{\mathbf x}_n) & \equiv
&
  \frac{1}{\beta} \, \int d\tau_1 \ldots d\tau_n \;
  {\mathcal W}^{(n)}_{a_1 \ldots a_n}(x_1,\,\ldots,\,x_n) \nonumber\\
  \times \; \big( \delta(\tau_1-\beta) - \delta(\tau_1) \big)
  &\ldots& \big( \delta(\tau_n-\beta) - \delta(\tau_n) \big) \nonumber \;.
\end{eqnarray}

Of course, one will usually evaluate just a few of the terms in the
expansion for ${\mathcal W}$, and as a consequence will obtain an
approximation to
the effective Hamiltonian.  From the first non-trivial order, namely,
keeping only the first term, the $2$-point correlation
function~\footnote{The meaning of this approximation shall be discussed at
the end of this calculation.} we see that:
\begin{equation}
  H_{eff}(\xi) \sim  H^{(2)}_{eff}(\xi) 
\end{equation}
where
\begin{equation}
  H^{(2)}_{eff}(\xi) \;\equiv\; \frac{1}{2} \, \int d{\mathbf x}_1
d{\mathbf x}_2
  \xi_a({\mathbf x}_1)\, H^{(2)}_{ab}({\mathbf x}_1,\,{\mathbf x}_2)
  \, \xi_b({\mathbf x}_2)
\end{equation}
and
\begin{eqnarray}\label{eq:Heffquad}
  H^{(2)}_{ab}({\mathbf x}_1,\,{\mathbf x}_2) &=& \frac{1}{\beta}
  \,\left[{\mathcal W}^{(2)}_{ab}(0,{\mathbf x}_1\,;0,
    {\mathbf x}_2)+ {\mathcal W}^{(2)}_{ab}(\beta ,{\mathbf x}_1\,;\beta ,
    {\mathbf x}_2)\right. \nonumber\\
  & & \left.  - {\mathcal W}^{(2)}_{ab}(\beta ,{\mathbf x}_1\,;0,{\mathbf
x}_2)-
    {\mathcal W}^{(2)}_{ab}(0,{\mathbf x}_1\,;\beta ,{\mathbf x}_2)
  \right] \;.
\end{eqnarray}
Introducing Fourier transforms for the correlation functions
$\widetilde{\mathcal W}_{ab}$ and the kernel, $\widetilde{H}^{(2)}_{ab}$,
(\ref{eq:Heffquad}) implies:
\begin{eqnarray}\label{eq:Heffquad1}
  \widetilde{H}^{(2)}_{ab}(\omega,{\mathbf k}) &=& \frac{2}{\beta}
  \,\int_{-\infty}^{+\infty} \frac{d\omega}{2\pi}\; 
  \big[ 1 \,-\, \cos(\beta \omega) \big] 
  \widetilde{\mathcal W}^{(2)}_{ab}(\omega,{\mathbf k})\;.
\end{eqnarray}

On the other hand, the kernels ${\mathcal W}^{(2)}_{ab}$ can be obtained
from the $T=0$ correlation function ${\mathcal W}^{(2)}_{11}$ which is,
essentially, the full quantum propagator for the $\varphi$-field:
\begin{equation} 
{\mathcal W}^{(2)}_{11}(x_1,x_2) \;=\; -\, \langle \varphi
  (x_1) \varphi(x_2) \rangle \;.
\end{equation}
Indeed, the explicit relations between the correlation functions ${\mathcal
  W}^{(2)}_{ab}$ with $a \neq 1$ or $b \neq 1$ and ${\mathcal
W}^{(2)}_{11}$,
in momentum space, are:
\begin{eqnarray}
  \widetilde{\mathcal W}^{(2)}_{22}(\omega,{\mathbf k}) 
  &=& \,1\,-\,\omega^2 \widetilde{\mathcal W}^{(2)}_{11}(\omega,{\mathbf
k}) 
  \nonumber\\
  \widetilde{\mathcal W}^{(2)}_{12}(\omega,{\mathbf k})&=& - \omega
\,\widetilde{\mathcal
    W}^{(2)}_{11}(\omega,{\mathbf k}) \nonumber\\
  \widetilde{\mathcal W}^{(2)}_{21}(\omega,{\mathbf k}) &=& 
\omega\widetilde{\mathcal
    W}^{(2)}_{11}(\omega,{\mathbf k})) \;.
\end{eqnarray}
In order to evaluate the frequency integrals, we need of course to make
some assumptions about the structure of $\widetilde{\mathcal
W}^{(2)}_{11}$.
Assuming that there is only one stable particle, and that the
renormalization conditions corresponding to the physical mass and
wavefunction renormalization
have been imposed. Then $\widetilde{\mathcal W}^{(2)}_{11}$ has only two
single poles, at the points $\pm \, i \, E_{ren}({\mathbf k})$, where
$E_{ren}({\mathbf k})$ is the dispersion relation, with a renormalized
mass, and including all the quantum corrections.

Considering each component of equation (\ref{eq:Heffquad1}) for specific
values for $a$ and $b$, we see that their respective $\omega$ integrals are
easily evaluated under the previous assumptions. The results are:
\begin{equation}\label{eq:Heffquad11}
  \widetilde{H}^{(2)}_{11}({\mathbf k}) \;=\; E^{-1}_{ren}({\mathbf k}) \, 
  \Big[n_B\big(E_{ren}({\mathbf k})\big) + 1 \Big]^{-1} \;,
\end{equation}
\begin{equation}\label{eq:Heffquad22}
  \widetilde{H}^{(2)}_{22}({\mathbf k}) \;=\; E_{ren}({\mathbf k}) \, 
  \Big[n_B\big(E_{ren}({\mathbf k})\big)+1 \Big]^{-1}\;,
\end{equation}
while \mbox{$\widetilde{H}^{(2)}_{12}=\widetilde{H}^{(2)}_{21} =0$}.
 
Since the functional integral has the same structure as the one for the
free case, to evaluate ${\mathcal Z}(\beta)$ in this approximation we just
need to modify the expression for the energy in the free field result:
${\mathcal Z}(\beta) \;\sim\; {\mathcal Z}^{(2)}(\beta)$ where
\begin{equation} {\mathcal Z}^{(2)}(\beta) \;=\; \prod_{\mathbf k} \big[
  n_B(E_{ren}({\mathbf k})) + 1\big]
\end{equation}
which yields the free energy
\begin{equation}
  F^{(2)}(\beta)\;=\; \frac{1}{\beta} \, \int \frac{d^dk}{(2\pi)^d} \, \ln
\big( 1  \,-\, e^{-\beta E_{ren}({\mathbf k})}\big) \;.
\end{equation}
This expression looks like the free energy for a free field, with the
renormalized energies instead of the free ones.  This approximation amounts
to considering the particle states as non interacting, after the
self-energy corrections have been taken into account (in the $2$-point,
$T=0$ correlation function). A possible way to understand this
approximation is by the large-$N$ limit, whereby correlation functions
involving more than $2$ points are suppressed.

An interesting approximation scheme, related to the functional expansion,
is given by the High-Temperature expansion. In terms of $H_{eff}$, we see
that the only place where $\beta$ appears is in the source $j$. Performing
an expansion of $j$ in powers of $\beta$:
\begin{equation}
  j_a(x) \;=\; \xi_a({\mathbf x}) \;\sum_{n=0}^\infty \,(-1)^n
\,\frac{\beta^n}{n!}  
  \delta^{(n)}(\tau) \;,  
\end{equation} 
and inserting this expansion into the expression for $H_{eff}$, we see that
the kernels corresponding to its functional expansion become:
\begin{eqnarray}
  H_{a_1 \cdots a_m}^{(m)}({\mathbf x}_1,\cdots,{\mathbf x}_m) &=&
\sum_{n_1=1,\cdots,n_m}^{\infty}
  \frac{\beta^{n_1+ \cdots + n_m -1}}{n_1! \cdots n_m!} \nonumber\\
  &\times & \big[\partial_{\tau_1} \cdots \partial_{\tau_m} {\mathcal
W}^{(m)}_{a_1\cdots a_m} 
  (x_1,\cdots,x_m)\big]_{\tau_i \to 0} \;.
\end{eqnarray}
When $\beta \to 0$, the leading term is linear in $\beta$ and only
$H^{(2)}$ survives:
\begin{equation}
  H_{a_1 a_2}^{(2)}({\mathbf x}_1,{\mathbf x}_2) \simeq 
  \beta^2 \big[\partial_{\tau_1}\partial_{\tau_2} {\mathcal W}^{(2)}_{a_1
a_2} 
  (x_1,x_2)\big]_{\tau_1, \tau_2 \to 0} \;.
\end{equation}

Keeping higher order $N$ terms in the expansion:
\begin{eqnarray}\label{eq:HeffexpansionN}
  H_{eff}^{(N)}(\xi) &=& \sum_{n=2}^{N}\frac{1}{n!}\int
  d^d{\mathbf x}_1 \cdots d^d{\mathbf x}_n\,H^{(n)}_{a_1 \ldots
    a_n}({\mathbf x}_1,\,\ldots,\,{\mathbf x}_n)\nonumber\\
  &\times& \xi_{a_1}({\mathbf x}_1) \cdots \xi_{a_n}({\mathbf x}_n) \;,
\end{eqnarray}
the partition function naturally admits a perturbative expansion, with the
quadratic kernel $H^{(2)}_{ab}$ determining the propagator, and the higher
order terms the (nonlocal) vertices.

The final example we consider is a massive Dirac field in $d+1$ spacetime
dimensions. The procedure will be essentially the same as for the real
scalar field, once the relevant kinematical differences are taken into
account. The action $S_0^f$ for the free case is given by $S_0^f =\int
d^{d+1}x \bar{\psi}(\not\!\partial + m)\psi$ where \mbox{$\not\!\partial=
  \gamma_{\mu}\partial_{\mu}$}, $\gamma_{\mu}^\dagger=\gamma_{\mu}$ and
$\{\gamma_{\mu},\gamma_{\nu}\}=2 \delta_{\mu\nu}$.

We then impose antiperiodicity conditions for both fields:
\begin{equation}
  \psi\left(\beta,{\mathbf x}\right) \,=\, -{\psi} 
  \left(0,{\mathbf x}\right)\;,\;\;
  \bar{\psi}\left(\beta,{\mathbf x}\right) \,=\, -\bar{\psi} 
  \left(0,{\mathbf x}\right)
\end{equation}
as constraints on the Grassmann fields. Those conditions lead to the
introduction of the two $\delta-$functions:
\begin{eqnarray} {\mathcal Z}_0^f(\beta) &=& \int \, {\mathcal
D}\psi{\mathcal
    D}\bar{\psi}\, \delta \big(\psi(\beta,{\mathbf x})+\psi(0,{\mathbf
    x})\big)\; \delta \big(\bar{\psi}(\beta,{\mathbf x})
  +\bar{\psi}(0,{\mathbf x})\big) \nonumber\\
  &\times& \exp \Big[-S_0^f(\bar{\psi},\psi)\Big] \;.
\end{eqnarray}
Since the Dirac action is of the first-order, the introduction of two
constraints, and two Lagrange multipliers, appears in an even more natural
way than for the previous case.  Those auxiliary fields, denoted by
$\chi(\mathbf
x)$ and $\bar\chi(\mathbf x)$ must be time-independent Grassmann spinors. 
The resulting expression for ${\mathcal Z}_0^f(\beta)$ is then
\begin{equation} 
{\mathcal Z}_0^f(\beta) \;=\; {\mathcal N}\int \,
{\mathcal
    D}\chi{\mathcal D}\bar{\chi}{\mathcal D}\psi{\mathcal D}\bar{\psi}\,
  e^{-S_0^f(\bar{\psi},\psi)+ i \, \int d^{d+1}x \,(\bar{\eta}\psi +
    \bar{\psi}\eta )},
\end{equation}
where $\eta$ and $\bar{\eta}$ are (Grassmann) sources depending on $\chi$
and $\bar{\chi}$ through the relations:
\begin{equation}
  \eta(x)\,=\,\chi({\mathbf x})\big[\delta(\tau -\beta)
  +\delta(\tau)\big]\;,\;\;\;\bar{\eta}(x)\,=\,\bar{\chi} ({\mathbf
    x})\big[\delta(\tau -\beta)+\delta(\tau)\big]\;.
\end{equation}

Integrating out $\psi, \bar{\psi}$, we arrive to:
\begin{equation} 
{\mathcal Z}_0^f(\beta)\;=\; \int \, {\mathcal
    D}\chi{\mathcal D}\bar{\chi}\, \exp \Big[-\beta
  H_{eff}\big(\bar{\chi},\chi \big)\Big]
\end{equation}
where
\begin{equation}
  H_{eff}\big(\bar{\chi},\chi \big)= \int d^{d}x \,\int d^{d}y \,
  \bar{\chi}({\mathbf x})H^{(2)}\big({\mathbf x},{\mathbf y}\big
)\chi({\mathbf y})
\end{equation}
with:
\begin{eqnarray}
  H^{(2)}\big({\mathbf x},{\mathbf y}\big )&=& \langle \mathbf x,0|
  (\not\!\partial + m)^{-1}|\mathbf y,0 \rangle + \langle\mathbf
x,\beta|(\not\!\partial 
  + m)^{-1}|\mathbf y,\beta \rangle \nonumber\\ 
  &+& \langle \mathbf x,0|(\not\!\partial + m)^{-1}|\mathbf y,\beta \rangle
  + \langle \mathbf x,\beta|(\not\!\partial + m)^{-1}|\mathbf y,0
\rangle\nonumber\\
  &=&\frac{1}{\beta}\,\Big[ 2\,S_f\big(0,\mathbf x - \mathbf y\big) 
  + S_f\big(\beta,\mathbf x - \mathbf y\big) 
  + S_f\big(-\beta,\mathbf x - \mathbf y\big)\Big].
\end{eqnarray}
In the last line, $S_f$, denotes the Dirac propagator.  A quite
straightforward calculation shows that
\begin{equation}
  H\big({\mathbf x},{\mathbf y}\big )=\frac{1}{\beta}\,
  \langle\mathbf x|\hat{u}(1 - \hat{n}_F)^{-1}|\mathbf y>
\end{equation}
where $\hat{n}_F \equiv \Big( 1+ e^{\beta \hat{n}}\Big)^{-1}$ is the
Fermi-Dirac distribution function, written in terms of with
\mbox{$\hat{h}$}, the energy operator (defined identically to its real
scalar field counterpart).  $\hat{u}$ is a unitary operator, defined as
\begin{equation}
  \hat{u} \;\equiv\; \frac{\hat{h}_D}{\hat{h}}\, , \;\; \hat{h}_D \equiv
  {\mathbf\gamma}\,\cdot\,{\mathbf \nabla} + m \;.
\end{equation}  

Then we verify that:
\begin{equation} 
{\mathcal Z}_0^f(\beta) \;=\;\det\hat{u}\;
{\det}^{-1}\big[
  (1-\hat{n}_F)\,\mathbf I\big] \;,
\end{equation}
($\mathbf I\,\equiv$ identity matrix in the representation of
Dirac's algebra)
\begin{equation} 
{\mathcal Z}_0^f(\beta) \;=\;\left\{\prod_{\vec{p}} \Big[
1 + e^{-\beta E(\vec{p})}\Big] \right\}^{r_d}
\end{equation}
with $E({\mathbf p})=\sqrt{{\mathbf p}^2 + m^2}$ and $r_d\,\equiv$
dimension of the representation (we have used the fact that $\det\hat{u} =
1$).

Again, the procedure has produced the right result for the partition
function, with a normal-ordered Hamiltonian. On the other hand, for a Dirac
field in a static external background corresponding to a minimally coupled
Abelian gauge field the $A_{0}=0$ gauge, we have
\begin{equation}\label{eq:defsDA}
  S^f (\bar{\psi},\psi, A)\;=\;\int d^{d+1}x \,
  \Big[ \bar{\psi}(x)\big(\not\!\partial + i\,e\, 
  {\mathbf\gamma}\cdot {\mathbf A} ({\mathbf x}) + m \big)\psi(x) \Big] \;.
\end{equation}
The assumed $\tau-$ independence allows us to carry on the derivation
described for the free case, with minor changes, arriving to the
expression:
\begin{eqnarray} 
{\mathcal Z}^f(\beta) \;&=&\;\det\hat{u}({\mathbf A})\,
  {\det}^{-1}\big(\hat{n}_{F}({\mathbf A})\,\mathbf I\big)\nonumber\\
  &=& e^{i K({\mathbf A})} \det\Big[\big( 1 + e^{-\beta \hat{h}({\mathbf
A})
  }\big)\mathbf I\Big]
\end{eqnarray}
where
\begin{equation}
  \hat{h}(A)\equiv \sqrt{-{\mathbf D}^2+m^2}\;,\;\;
  {\mathbf D}\equiv {\mathbf\nabla}-i e {\mathbf A} \;,
\end{equation}
and:
\begin{equation}
  e^{i K({\mathbf A})}\;=\;\frac{\det\big( {\mathbf\gamma}\cdot {\mathbf D}
+ m \big)}{\det   \sqrt{-{\mathbf D}^{2} + m^{2}}}\;.
\end{equation}
Notice that the factor \mbox{$\det\Big[\big( 1 + e^{-\beta \hat{n}({\mathbf
   A})}\big) \mathbf I\Big]$} can be formally diagonalized in terms of
the energies $E_\lambda({\mathbf A})$ in the presence of the external
field. Thus we arrive to the expression:
\begin{equation} 
{\mathcal Z}^f(\beta) \;=\;e^{i K({\mathbf A})}\;\times \;
  \left\{\prod_\lambda \Big[ 1 + e^{-\beta E_\lambda({\mathbf A})}\Big]
  \right\}^{r_d} \;.
\end{equation}
The factor $e^{i K({\mathbf A})}$, on the other hand, is topological in
origin, as it depends on the phase, $K({\mathbf A})$, of the determinant of
$\hat{h}_D$. On the other hand, $\hat{h}_D$ may be regarded as a kinetic
operator in one fewer dimension.  For Dirac fermions, we know that the
phase of $\det\hat{h}_D$ can be non-trivial only when $d$ is odd, i.e.,
when $d+1$ is even. However, the $\gamma$-matrices appearing in
$\det\hat{h}_D$ form a reducible representation of the Dirac algebra in $d$
dimensions, with the matrix $\gamma_\tau$ relating every eigenvalue to its
complex conjugate. Thus, as a result, the phase $K({\mathbf A})$ vanishes.
Of course, a non-vanishing result may be obtained for other fermionic
systems, like Weyl fermions for \mbox{$d+1={\rm even}$}.

Summarizing, we have shown that, by introducing the `thermal'
boundary conditions as constraints in the Euclidean path integral for the
vacuum functional, we may obtain a novel representation for the partition
function.
This representation may be thought of as an integral over the phase space
variables, weighted by a Boltzmann factor corresponding to an effective
quantum Hamiltonian, $H_{eff}$, which reduces to the classical one in the
corresponding (high-temperature) limit.  We discussed the main properties
of this representation for the cases of the real scalar and Dirac fields,
two typical examples that have been chosen for the sake of simplicity. It
is not difficult to generalize the representation to the case of systems
containing fermions interacting with bosons, by introducing the proper
$T=0$ generating functionals.

\section*{Acknowledgements}
C.D.F. thanks CONICET (PIP5531) and CAPES/SPU for financial support.
A.P.C.M.
and I.R.  thanks CAPES/SPU, CNPq/MCT and Pronex/FAPERJ for partial
financial
support.

\end{document}